# SCKF-LSTM Based Trajectory Tracking for Electricity-Gas Integrated Energy System

Liang Chen, Yang Li, *Senior Member, IEEE*, Jun Cai, Songlin Gu, and Ying Yan

*Abstract*—A novel approach of tracking the dynamic trajectories for electricity-gas interconnected networks is developed in the studies, leveraging a Kalman filter-based structure. To capture the accurate system trajectories, the Holt's exponential smoothing techniques and nonlinear dynamic equations of gas pipelines are applied to establish the power and gas system equations, respectively. Addressing the numerical challenges posed by the strongly nonlinear system, a square-root cubature Kalman technique based tracking solution is adopted. For the effectiveness in time series prediction, the mass flow rates forecasting task of gas loads is undertaken by employing a long short-term memory network at each computation step. Consequently, a combined method for tracking the dynamic trajectories of comprehensive energy systems by combining these two algorithms is constructed. The IEEE-39 bus network as well as the GasLib-40 node gas network is integrated by gas turbine units to form the multi-energy network, and two indexes are introduced for a numerical analysis of the tracking performances. The outcomes demonstrate that the suggested approach significantly improves tracking accuracy when contrasted with the reference measurements.

This work was supported in part by the National Natural Science Foundation of China under Grant 52077105, and in part by Excellent Research and Innovation Teams in Universities in Anhui Province under Grant 2023AH010021 (Corresponding author: Liang Chen, Yang Li and Jun Cai.)

Liang Chen is with C-MEIC, CICAEET, School of Automation, Nanjing University of Information Science and Technology, Nanjing, 210044, China (e-mail: ch.lg@nuist.edu.cn).

Yang Li is with the School of Electrical Engineering, Northeast Electric Power University, Jilin 132012, China (e-mail: liyang@neepu.edu.cn).

Jun Cai, is with C-MEIC, CICAEET, and School of Automation, Nanjing University of Information Science and Technology, Nanjing 210044, China; he is also with School of Mechanical and Electrical Engineering, Anhui Jianzhu University, Hefei 230009, China. (E-mail: j.cai@nuist.edu.cn)

Songlin Gu is with State Grid Economic and Technological Research Institute Co., Ltd, Beijing, 102209, China (e-mail: gsl0516@163.com).

Ying Yan is with C-MEIC, CICAEET, School of Automation, Nanjing University of Information Science and Technology, Nanjing, 210044, China; he is also with School of Mechanical and Electrical Engineering, Anhui Jianzhu University, Hefei, 230009 (e-mail: ying.yan@nuist.edu.cn).

## I. INTRODUCTION

WITH the escalating demand for sustainable and efficient energy in our increasingly eco-conscious world, integrated energy systems (IESs) [1] emerge as a pivotal solution to contemporary energy challenges. IESs epitomize the synergy of various energy resources, orchestrating them into a cohesive unit for a more efficient, clean, and reliable energy supply. The significance of IESs extends beyond mere energy sustainability; they are instrumental in enhancing the overall efficiency of energy systems, curtailing environmental pollution, and fostering sustainable development.

However, the efficient operation and management of IESs have many challenges [2]. Due to the involvement of various energies in the system, such as electric power, nature gas pipelines and heating networks, the complex interactions and dynamic processes between these energy systems make system optimal operation [3] and control [4] very complex. To ensure efficiency, it is crucial to accurately track the real-time states of the IESs, forecast future states, and make informed decisions for intelligent scheduling and optimization.

Traditionally, least squares-based state estimation [5] has been employed to model and process data, providing precise static state information vital for optimal operation and control. However, with the advent of new energy sources, the dynamics of IESs have grown increasingly complex. As a result, conventional static state estimation methods may no longer suffice for the real-time monitoring and optimization required in modern IESs.

An effective solution to the aforementioned problem is the implementation of a tracking method based on Kalman filter structure [6], a widely-used mathematical technique for estimating the states of dynamic systems. It combines the system dynamic model with real-time observed data to provide accurate dynamic state information [7]. The fundamental principle involves integrating the system dynamic model and measurements to minimize the expected error of the estimation. To tracking the dynamics of the generators, the unscented Kalman filter (UKF) is applied in [8]. The principle is combining generator dynamic equations and measurements to obtain high accurate dynamic trajectory. As for the IESs, the dynamic models should be established firstly. In [9], the nonlinear dynamic equations of gas networks are linearized based on some given assumptions, and then the linear model of gas systems is proposed, which makes it possible to track the dynamic trajectory of the IESs by using Kalman filter method.

# Nomenclature

| | | | |
|---|---|---|---|
| $\alpha_E, \beta_E$ | smoothing indexes, $\alpha_E, \beta_E \in [0,1]$ | $\mathbf{v}, \mathbf{w}$ | predicting errors and measuring errors |
| $\mathbf{x}_{Et}$ | state vector of electric power systems | $\mathbf{S}_Q, \mathbf{S}_R$ | covariance of predictions and measurements |
| $n_B$ | the total number of buses | $\widehat{\mathbf{X}}_{i,t}$ | cubature point of tracking values |
| $e_{it}, f_{it}$ | voltage phasor real part and imaginary parts | $\widehat{\mathbf{S}}_t$ | tracking error covariance |
| $\rho, \phi$ | gas density and gas flow rate | $n_x$ | total number of states |
| $\tau, \varsigma$ | time and space along pipelines | $\widehat{\mathbf{x}}_t$ | tracking state vector |
| $d, A$ | diameter and cross section area of pipelines | $\widetilde{\mathbf{X}}^*_{i,t+1}$ | predicted cubature point |
| $c_S$ | constant sound speed | $\widetilde{\mathbf{x}}_{t+1}$ | predicted states |
| $\gamma$ | friction factor | $\widetilde{\mathcal{X}}^*_{t+1}$ | centered matrix of cubature points |
| $L_{Pij}, d_{ij}, A_{ij}$ | length, diameter and cross section area of pipeline $ij$ | $\widetilde{\mathbf{S}}_{t+1}$ | predicted error covariance |
| $\Delta t$ | time step | $\widetilde{\mathbf{X}}_{i,t+1}$ | cubature point of predicted states |
| $n_P$ | total number of pipelines | $\widetilde{\mathbf{Z}}_{i,t+1}$ | predicted cubature point of measurements |
| $\rho_{cs}$ | constant density of source node $s$ | $\widetilde{z}_{t+1}$ | predicted value of measurements |
| $n_G$ | total number of nodes | $\widetilde{\mathcal{Z}}_{t+1}$ | centered matrix of measurements |
| $\phi_{l,t+1}$ | gas load at node $l$ | $\mathbf{S}_{zz,t+1}$ | error covariance of measurements |
| $P_{Gi,t+1}$ | output power at time $t+1$ of the GTU at bus $i$ | $\widetilde{\mathcal{X}}_{t+1}$ | centered matrix of states |
| $\eta_i$ | energy conversion coefficient | $\mathbf{P}_{xz,t+1}$ | cross-covariance matrix |
| $z^E_{b,t}, z^F_{b,t}$ | voltage measurements of bus $b$ | $\mathbf{W}_{t+1}$ | weighted matrix |
| $z^{BR}_{ij,t}, z^{BI}_{ij,t}$ | currents measurements of branch $ij$ | $\widehat{\mathbf{x}}_t, \widetilde{\mathbf{x}}_{t+1}$ | tracking states and predicted states |
| $z^{IR}_{b,t}, z^{II}_{b,t}$ | injected currents measurements of bus $b$ | $\sigma$ | sigmoid activation function |
| $g_{ij}, b_{ij}$ | branch conductance and susceptance of transmission line $ij$ | tanh | hyperbolic tangent activation function |
| $g_{i0}, b_{i0}$ | shunt conductance and susceptance at bus $i$ | $f_t, i_t, o_t$ | outputs of the 3 gates in LSTM network |
| $G_{bi}, B_{bi}$ | elements of row $b$ and column $i$ in the admittance matrix | $W_i, \kappa_i, \varepsilon_i$ | network parameters of input gate |
| $z^P_{n,t}, z^M_{n,t}$ | measurements of pressures and mass flows at gas node $n$ | $W_f, \kappa_f, \varepsilon_f$ | network parameters of forget gate |
| $\mathbf{f}, \mathbf{h}$ | system equations and measurement equations | $W_o, \kappa_o, \varepsilon_o$ | network parameters of output gate |
| $\mathbf{x}_{t+1}, \mathbf{z}_{t+1}$ | state and measurement vector at time $t+1$ | $X_{Lt}, \omega_t$ | input and output of LSTM |
| $\mathbf{z}_{Et+1}, \mathbf{z}_{Gt+1}$ | measurement vectors of power and gas systems | $\otimes$ | vector production |
| $\mathbf{z}^E_{t+1}, \mathbf{z}^F_{t+1}$ | measurement vectors of voltage real and imaginary parts | $M_{LN}$ | normalized load |
| $\mathbf{z}^{BR}_{t+1}, \mathbf{z}^{BI}_{t+1}$ | measurement vectors of branch current real and imaginary parts | $M_{Lmin}, M_{Lmax}$ | minimal and maximal load value |
| $\mathbf{z}^{IR}_{t+1}, \mathbf{z}^{II}_{t+1}$ | measurement vectors of injected current real and imaginary parts | $\mathbf{C}, \mathbf{\Phi}$ | Jacobian matrix of system equation and measurement equation |
| $\mathbf{z}^P_{t+1}, \mathbf{z}^M_{t+1}$ | pressure and mass flow measurements | | |

Based on the linear model, a trajectory tracking method for electricity, gas and heating systems is proposed in [10]. Further, to deal with the bad data in the measurements, the multidimensional scaling factor is introduced to regulate the measurement weights, and a robust dynamic tracking method of IESs based on Kalman filter is proposed in [11]. However, it's important to note that most practical systems are nonlinear, and the linearization could lead to significant errors in the dynamic models.

Aiming at the above problem, the nonlinear equations are linearized, and the extended Kalman filter (EKF) [12] algorithm is formulated. Additionally, by representing the posterior probability of a random event through a set of weighted random samples, known as particles, the particle filter (PF) is applied in [13]. Further, based on the similar idea, the UKF [14] are extensively applied to deal with nonlinear problems. In [15], the cubature Kalman fiter (CKF) is generated, which approximates the integral of the Gaussian distribution through a set of specific points with the same weight, known as cubature points. To enhance the robustness of UKFs, a novel robust adaptive method is proposed in [16], which is able to detect and identify gross errors in measurements. In [17], to enhance the robustness of CKF, a robust estimation algorithm is proposed. To obtain the accurate charging states, a battery state estimator is proposed in [18] and [19], by using the square-root cubature Kalman filter (SCKF). Confronting the challenge of noise interference, the research documented in reference [20] introduces an adaptive technique and improves the SCKF for assessing battery charging status. With the developments of artificial intelligence technologies, many combined method appears [21]. To achieve the high prediction and training efficiencies, the convolutional neural network (CNN) is combined with CKF in [22] for positioning. To deal with the uncertainty, an interval probability distribution learning model and rough auto-encoder model based wind speed predicting method is proposed in [23] and [24], respectively. In [25], to enhance the estimator's precision, the long short term memory (LSTM) network is combined with CKF to form an integrated approach for estimating the charging states. In [26], a fusion algorithm based on locally weighted linear regression, LSTM and CKF is proposed for the navigation systems, which is able to estimate the attitude accurately. It can be seen that the CKF is a powerful tool for filtering and tracking problems, while the integrated method of CKF and artificial intelligence algorithms track many researchers' interests. However, the integrated methods for the dynamic trajectory tracking problem have not been covered yet.

In our studies, a novel method for tracking the IESs' states is developed. The unique novelties of the method compared with existed studies are shown in Tab.1, and the primary contributions are outlined below:

1) Establishment of Detailed Trajectory Tracking Models for IESs: For natural gas pipelines, dynamic processes involving gas pressures and mass flow rates are modeled as nonlinear PDEs. These PDEs are transformed into difference equations using the Euler method for predicting the states of the gas system. In power systems, due to the challenges in establishing dynamic models for voltage phasors, Holt's exponential smoothing technology is employed for state prediction. This leads to the formulation of comprehensive nonlinear system models for IESs by integrating PDEs with Holt's exponential smoothing technology. Measurement equations for Phasor Measurement Units (PMUs), as well as gas pressures and mass flow rates, are also established.

2) Proposal of an SCKF Structure-Based Trajectory Tracking Method: The intense nonlinearity and ill-conditioning of the system models render the classical CKF algorithm ineffective for these tracking models. To tackle this issue, the triangular square-root factor is computed instead of matrix triangular factorizations, leading to the development of a tracking method based on the SCKF structure.

3) Development of an SCKF-LSTM Based Trajectory Tracking Method for IESs: The prediction of gas loads is facilitated by the LSTM network. The gas loads in the nonlinear system model of the gas system have to be predicted at each time steps. The combination of LSTM network and SCKF result in the proposal of the SCKF-LSTM based trajectory tracking method for IESs.

TABLE I
A COMPARISON OF EXISTING METHODS WITH THE PROPOSED METHOD

| Ref. | static/dynamic | model | algorithm |
|---|---|---|---|
| [27] | static | nonlinear | particle swarm optimization |
| [28] | static | nonlinear | WLS |
| [10],[11], [29], [30] | dynamic | linear | Kalman filter |
| this study | dynamic | nonlinear | SCKF-LSTM |

Keeping the above perspective, the following sections provide an in-depth look at the innovative tracking approach that utilizes the SCKF-LSTM algorithm and the outcomes derived from applying this method in a demonstrative scenario. Section II outlines the construction of the comprehensive tracking models for IESs. Section III presents the trajectory tracking methodology that is grounded in the SCKF-LSTM framework. Section IV details the case study analysis. Conclusive remarks are encapsulated in Section V.

## II. MODELING OF ELECTRICITY-GAS NETWORKS

Precise tracking of state trajectories is crucial for the efficient management and regulation of IESs. A potent approach for state tracking is the method utilizing a Kalman filter-structured trajectory tracking framework.

### A. Basic Idea of the Trajectory Tracking

The basic process of Kalman filter based tracking method includes prediction and filtering steps. Figure 1 illustrates the fundamental framework for the novel trajectory tracking method. The electric power system states are denoted by $x_{Et}$, and exponential smoothing techniques are applied to predict the states. The vector $x_{Gt}$ is gas states, which are forecasted by using the PDEs. The accurate tracking states $\hat{x}_{t+1}$ are obtained by balancing measurements $z_{t+1}$ and the predicting value of states $\tilde{x}_{t+1}$ in the filtering step.

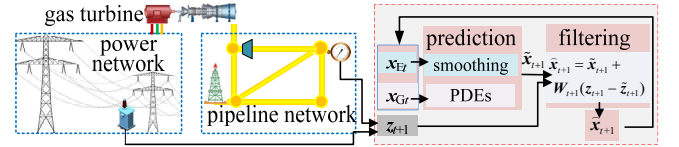

Fig. 1. The structure of the tracking method.

### B. Holt's Exponential Smoothing Technique

The changes of bus voltages in dynamic processes are affected by the fluctuations of power loads, the randomness of which is strong; as a result, it is impractical to construct precise dynamic models for electric power systems. As a result, numerical forecasting methods have to be applied in the method. The universally applied Holt's exponential smoothing method enables the forecasting of stochastic time series, akin to the fluctuations of voltages. The Holt's exponential smoothing method is shown as follows:

$$s_{Lt+1} = \alpha_E x_{Et} + (1-\alpha_E)(s_{Lt}+s_{Tt}) \qquad (1)$$
$$s_{Tt+1} = \beta_E(s_{Lt+1}-s_{Lt})+ (1-\beta_E) s_{Tt} \qquad (2)$$
$$x_{Et+1} = s_{Lt} + s_{Tt} \qquad (3)$$

where, $x_{Et} = [e_{1t}, f_{1t}, e_{2t}, f_{2t}, \ldots, e_{n_B t}, f_{n_B t}]^T$, $x_{Et} \in \mathbb{R}^{2n_B \times 1}$. The initialization of $s_{Lt}$ is $x_{E2}$, and $x_{E2}-x_{E1}$ for $s_{Tt}$. By substituting (1) and (2) into (3), the following system equations of power systems are obtained:

$$x_{Et+1} = \alpha_E x_{Et} + (1-\alpha_E)(s_{Lt-1}+s_{Tt-1}) + s_{Tt} \qquad (4)$$

### C. Dynamic Model of Gas Systems

The pressures at different locations along pipelines are influenced by the changes of gas flows due to compressions. This dynamic behavior is typically characterized and modeled using PDEs. In pipelines where gas flow speed significantly falls below the sound speed, the convective influences of the gas may be considered negligible. Therefore, for horizontal gas pipelines, the dynamic equations are as follows [9]:

$$\frac{\partial \rho}{\partial \tau} + \frac{\partial \phi}{A \partial \varsigma} = 0 \qquad (5)$$

$$\frac{\partial \phi}{A \partial \tau} + c_S^2 \frac{\partial \rho}{\partial \varsigma} + \frac{\gamma \phi^2}{2dA^2 \rho} = 0 \qquad (6)$$

In the context of natural gas networks, the intersections of multiple pipelines are designated as nodes. The distance of an individual pipeline segment connecting node $i$ with node $j$ is denoted as $L_{Pij}$, while the densities are $\rho_i$ and $\rho_j$, respectively. The detailed modeling method for a single pipe is shown as Fig. 1. The variables $\phi_i$ and $\phi_j$ represent the gas loads, and the subscript $i$ and $j$ are node number. The mass flow rates in pipeline $ij$ are $\phi_{ij}$ and $\phi_{ji}$, respectively. The green trapezoids are compressors, and the density ratios are $c_{rij}$ and $c_{rji}$, respectively.

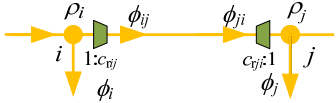

Fig. 2. The model of gas pipelines.

The state dynamics of a single pipeline satisfy (5) and (6), which have to be differenced by Euler finite difference technique before solving. For the pipeline model in Fig. 2, the densities and masses can be determined through the use of the following difference equations:

$$L_{Pij}A_{ij}c_{r,ij}(\rho_{i,t+1}-\rho_{i,t})+ L_{Pij}A_{ij}c_{r,ji}(\rho_{j,t+1}-\rho_{j,t}) + (\phi_{ji,t+1}- \phi_{ij,t+1})\Delta t + (\phi_{ji,t} - \phi_{ij,t})\Delta t = 0 \qquad (7)$$

$$\phi_{ij,t+1} - \phi_{ij,t} + \phi_{ji,t+1} - \phi_{ji,t} + A_{ij}\Delta t c_S(c_{r,ji}\rho_{j,t+1} - c_{r,ij}\rho_{i,t+1}) + A_{ij}\Delta t c_S(c_{r,ji}\rho_{j,t} - c_{r,ij}\rho_{i,t})$$
$$+ \frac{\Delta t L_{Pij} \gamma_{ij} \left(\phi_{ij,t}+\phi_{ji,t}\right)^2}{4 d_{ij} A_{ij}(c_{rij}\rho_{i,t} + c_{rji}\rho_{j,t})} = 0 \qquad (8)$$

The state vector of gas systems is $x_{Gt} = [\rho_{1,t}, \cdots, \rho_{n_G,t}, \cdots, \phi_{ij,t}, \phi_{ji,t}, \cdots]^T$, $x_{Gt} \in \mathbb{R}^{n_G+2n_P \times 1}$. The terms of time $t$ in (7) and (8) are moved to the right side, and then (9) and (10) are obtained:

$$A_{ij}L_{Pij}(c_{rij}\rho_{i,t+1} + c_{rji}\rho_{j,t+1}) + \Delta t(\phi_{ji,t+1}- \phi_{ij,t+1})$$
$$= L_{Pij}A_{ij}(c_{rij}\rho_{i,t} + c_{rji}\rho_{j,t}) + (\phi_{ij,t}- \phi_{ji,t})\Delta t \qquad (9)$$

$$L_{Pij}(\phi_{ji,t+1}+\phi_{ij,t+1}) + A_{ij}\Delta t c_S\left(c_{rji}\rho_{j,t+1} - c_{rij}\rho_{i,t+1}\right)$$
$$= L_{Pij}(\phi_{ji,t}+\phi_{ij,t}) + A_{ij}\Delta t c_S\left(c_{rij}\rho_{i,t} - c_{rji}\rho_{j,t}\right) \qquad (10)$$
$$- \frac{L_{Pij}\gamma_{ij}\left(\phi_{ji,t}+\phi_{ij,t}\right)^2 \Delta t}{4 d_{ij} A_{ij}(c_{rij}\rho_{i,t} + c_{rji}\rho_{j,t})}$$

For every single pipeline, the densities and flows should satisfy (9) and (10). The objective of establishing (9) and (10) is to calculate the gas states. It can be seen that there are 4 states for each pipeline, but only 2 equations are established, therefore, more equations have to be given. Firstly, all nodes in gas networks are classified into 2 categories: source and load. The source nodes connect gas sources, and the gas densities are constant. The load nodes connect gas loads, and the node mass flow balance should be satisfied. The above two constraints are called boundary conditions, and the detailed models are:

$$\rho_{s,t+1} = \rho_{Cs}, \; s\text{: source node}, s=1, 2, \ldots, S \qquad (11)$$

$$\sum_{j\in m}\phi_{mj,t+1} = \phi_{m,t+1}, \; m\text{: load node}, l= S+1, S+2, \ldots, n_G \qquad (12)$$

where, $j \in m$ represents node $j$ connects node $m$. In the IESs, the two kinds of networks are coupled though gas turbine units (GTUs), and the gas is converted to electric powers. Gas loads are related to the generated electric power of GTUs, and the detailed model is:

$$\phi_{m,t+1} = \frac{P_{Gi,t+1}}{\eta_i}, \; i \triangleleft m \qquad (13)$$

where, $i \triangleleft m$ means that gas node $m$ and bus $i$ are connected through the GTU. The matrix form of (9)-(12) is

$$A_G x_{Gt+1} = f_G(x_{Gt}, u_{t+1}) \qquad (14)$$

where, $u_{t+1} = [\rho_{C1}, \rho_{C2}, \ldots, \rho_{CS}, \phi_{S+1,t+1}, \ldots, \phi_{n_G,t+1}]^T$; $A_G$ is the coefficient matrix of state vector $x_{Gt+1}$; $f_G$ is nonlinear functions. The nonlinear dynamic model of gas systems can be obtained:

$$x_{Gt+1} = A_G^{-1} f_G(x_{Gt}, u_{t+1}) \qquad (15)$$

By using the dynamic model (4) and (15), the states of IESs can be predicted. In Kalman filter based tracking method, in addition to dynamic models, the measurement equations should be given also.

### D. Measurement Equations

In the dynamic tracking problems of IESs, all of the measurements should be at the same time, which means that the measurements must be synchronous. With the help of PMUs, the data acquisitions can satisfy the above requirement by using the satellite synchronous clock [31]. The PMUs are able to measure power system voltage and current phasors directly, which are taken as measurements of power systems. The measurement equations of power systems are:

$$z_{b,t}^E = e_{bt} \qquad (16)$$
$$z_{b,t}^F = f_{bt} \qquad (17)$$
$$z_{ij,t}^{BR} = (g_{ij} + g_{i0})e_i - (b_{ij} + b_{i0})f_i - g_{ij}e_j + b_{ij}f_j \qquad (18)$$
$$z_{ij,t}^{BI} = (b_{ij} + b_{i0})e_i + (g_{ij} + g_{i0})f_i - b_{ij}e_j - g_{ij}f_j \qquad (19)$$
$$z_{b,t}^{IR} = \sum_i (G_{bi}e_i - B_{bi}f_i), \; i \rangle b \qquad (20)$$
$$z_{b,t}^{II} = \sum_i (B_{bi}e_i + G_{bi}f_i), \; i \rangle b \qquad (21)$$

The measurement equations of gas systems are:
$$z_{n,t}^{P} = c_S^2 \rho_{n,t} \quad (22)$$
$$z_{n,t}^{M} = \sum_i \phi_{in,t}, \ i \in n \quad (23)$$

## III. SCKF-LSTM Based Trajectory Tracking Method

In the CKF-based tracking method, state prediction relies on utilizing dynamic models. Crucially, before these predictions can be made, the gas loads must first be estimated using the tracking values generated by LSTM. The paper elaborates on the detailed structure of the SCKF-LSTM tracking method, providing a comprehensive overview of its implementation and functionality.

### A. The Structure of SCKF-LSTM Tracking

The tracking approach utilizing the CKF involves two principal phases: the prediction phase and the filtering phase [32]. For the prediction phase, the dynamic models are applied to forecast the state vector $\tilde{x}_{t+1}$ by tracking state vector $\hat{x}_t$; while in the filtering phase, $\tilde{x}_{t+1}$ is corrected using measurements. First of all, the dynamic models and measurement equations should be given:
$$\begin{cases} x_{t+1} = f(x_t, u_{t+1}) + v \\ z_{t+1} = h(x_{t+1}) + w \end{cases} \quad (24)$$

where, $f$ is composed of (4) and (15); $h$ is composed of (16)-(23); $x_{t+1} = [x_{Et+1}^T \ x_{Gt+1}^T]^T$; $z_{t+1} = [z_{Et+1}^T \ z_{Gt+1}^T]^T$; $z_{Gt+1} = [(z_{t+1}^P)^T (z_{t+1}^M)^T]^T$; $z_{Et+1} = [(z_{t+1}^E)^T (z_{t+1}^F)^T (z_{t+1}^{BR})^T (z_{t+1}^{BI})^T (z_{t+1}^{IR})^T (z_{t+1}^{II})^T]^T$.

The dynamic models in (24) are high-dimensional strong nonlinear equations, and the CKF cannot deal with the tracking problem, as a result, the SCKF algorithm based tracking method is proposed. In the method, all of the mass flow rates at sink nodes should be obtained in each step. However, it is impossible to install flowmeters at all nodes in the gas systems. To deal with this problem, the tracking values of mass flow rates are applied to predict the node mass flow rates by using LSTM, and the SCKF-LSTM based tracking method is proposed, and the structure is given in Fig.3. At time step $t+1$, the tracking states $\hat{x}_t$ at time $t$, measurements $z_{t+1}$ and variables $u_{t+1}$ including gas source densities $\rho_{Cs}$ and the predicted load flow at sink nodes $\tilde{\phi}_{l,t+1}$ are the inputs of the SCKF. The tracking values of gas load $\hat{\phi}_{l,t+1}$ are used to predict $\tilde{\phi}_{l,t+2}$ by the SLTM. The tracking states $\hat{x}_{t+1}$ and $\tilde{\phi}_{l,t+2}$ are the outputs of the tracking methods. The tracking values of state $\hat{x}_t$ is applied to forecast the states in SCKF for the next step, while the tracking values of mass flow rates $\hat{\phi}_{ij,t}$ in $\hat{x}_t$ are transformed to mass flow of gas load $\hat{\phi}_{l,t}$ for predicting $\tilde{\phi}_{l,t+1}$ by the LSTM method. If the tracking step is $\Delta t$, the time window for LSTM is $w \cdot \Delta t$. In addition to $\tilde{\phi}_{l,t+1}$, the gas densities $\rho_{Cs}$ at source nodes are the other elements in variable $u_{t+1}$, which are constant values in the tracking processes.

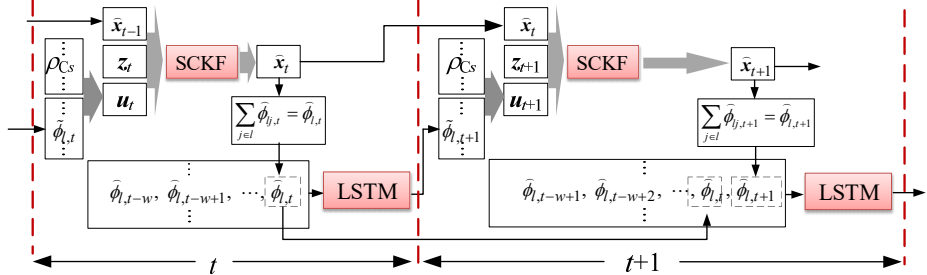

Fig. 3. The structure of SCKF-SLTM based tracking method.

### B. The SCKF Algorithm

Addressing the challenge of high-dimensional, strongly nonlinear tracking problems (24), the traditional CKF encounters limitations because of the non-positive definiteness of covariance matrix. To counter this issue, the SCKF is introduced in this study as a more robust solution for the tracking problem. The following sections detail the specific processes of the SCKF based mehtod [15]:

**Prediction step**

In this step, the predicted state $\tilde{x}_{t+1}$ and error covariance $\tilde{S}_{t+1}$ are generated:
$$\hat{X}_{i,t} = \hat{S}_t \, \varepsilon_{Ei} + \hat{x}_t, \ i = 1, 2, \ldots, 2n_x \quad (25)$$
$$\tilde{X}_{i,t+1}^* = f(u_{t+1}, \hat{X}_{i,t}) \quad (26)$$
$$\tilde{x}_{t+1} = \frac{1}{2n_x} \sum_{i=1}^{2n_x} \tilde{X}_{i,t+1}^* \quad (27)$$
$$\tilde{\mathcal{X}}_{t+1}^* = \frac{1}{\sqrt{2n_x}}[\tilde{X}_{1,t+1}^* - \tilde{x}_{t+1}, \ \cdots, \ \tilde{X}_{2n_x,t+1}^* - \tilde{x}_{t+1}] \quad (28)$$
$$\tilde{S}_{t+1} = \text{Tria}([\tilde{\mathcal{X}}_{t+1}^* \ S_Q]) \quad (29)$$

**Filtering step**

In this step, the tracking value $\hat{x}_{t+1|}$ and error covariance $\hat{S}_{t+1}$ are obtained:
$$\tilde{X}_{i,t+1} = \tilde{x}_{t+1} + \tilde{S}_{t+1} \, \varepsilon_{Ei}, \ i = 1, 2, \ldots, 2n_x \quad (30)$$
$$\tilde{Z}_{i,t+1} = h(\tilde{X}_{i,t+1}) \quad (31)$$
$$\tilde{z}_{t+1} = \frac{1}{2n_x} \sum_{i=1}^{2n_x} \tilde{Z}_{i,t+1} \quad (32)$$
$$\tilde{\mathcal{Z}}_{t+1} = \frac{1}{\sqrt{2n_x}}[\tilde{Z}_{1,t+1} - \tilde{z}_{t+1}, \cdots, \tilde{Z}_{2n_x,t+1} - \tilde{z}_{t+1}] \quad (33)$$
$$S_{zz,t+1} = \text{Tria}([\tilde{\mathcal{Z}}_{t+1} \ S_R]) \quad (34)$$

$$\tilde{\mathcal{X}}_{t+1} = \frac{1}{\sqrt{2n_x}}[\tilde{X}_{1,t+1} - \tilde{x}_{t+1}, \cdots, \tilde{X}_{2n_x,t+1} - \tilde{x}_{t+1}] \quad (35)$$

$$P_{xz,t+1} = \tilde{\mathcal{X}}_{t+1} \tilde{\mathcal{Z}}_{t+1}^T \quad (36)$$

$$W_{t+1} = (P_{xz,t+1}/S_{zz,t+1}^T)/S_{zz,t+1} \quad (37)$$

$$\hat{x}_{t+1} = \tilde{x}_{t+1} + W_{t+1}(z_{t+1} - \tilde{z}_{t+1}) \quad (38)$$

$$\hat{S}_{t+1} = \text{Tria}([\tilde{\mathcal{X}}_{t+1} - W_{t+1}\tilde{\mathcal{Z}}_{t+1} \; W_{t+1}S_R]) \quad (39)$$

where, $\varepsilon_{Ei}$ is the $i$th column vector of matrix

$$\varepsilon_E = \begin{pmatrix} 1 & 0 & \cdots & 0 & -1 & 0 & \cdots & 0 \\ 0 & 1 & \cdots & 0 & 0 & -1 & \cdots & 0 \\ \vdots & \vdots & \cdots & \vdots & \vdots & \vdots & \cdots & \vdots \\ 0 & 0 & \cdots & 1 & 0 & 0 & \cdots & -1 \end{pmatrix}, \; i = 1, 2, \ldots, 2n_x;$$

In the above processes, the LSTM net is used to predict gas loads, which are given in the next section.

### C. LSTM

The recent surge in artificial intelligence research has led to significant breakthroughs in various domains, notably in natural language processing and time series analysis. Central to these developments is the LSTM structure, a specialized kind of recurrent neural network (RNN) has gained considerable attention because it is able to obtain and model intricate temporal dependencies.

The LSTM network, a variant of the RNN, is specifically engineered to overcome the challenge of vanishing gradients, a prevalent challenge encountered in standard RNNs. The issue of vanishing gradients arises when the gradients diminish to minuscule values during the training phase, impeding the network's capacity to discern extensive temporal dependencies within sequential datasets. LSTMs deal with this problem by integrating an intricate gating mechanism.

LSTM networks are composed of memory cells, each featuring three principal regulatory mechanisms: an input gate, a forget gate, and an output gate [33], which empower LSTM cells to control the passage of data, enabling the network to intentionally retain or omit information from previous points in a sequence. The structure also includes a cell state that serves as an internal memory. The combination of these components grants LSTM the capacity to store and retrieve information over extended sequences, making it highly effective in sequential data modeling. The comprehensive structure of LSTM is illustrated in Fig. 4.

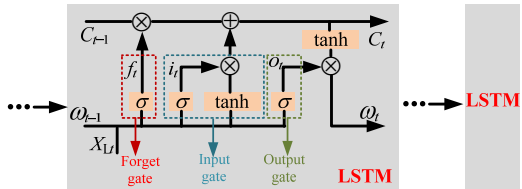

Fig. 4. The structure of SLTM.

The outputs of LSTM are computed by the following equations:

$$f_t = \sigma(W_f X_{Lt} + \kappa_f \omega_{t-1} + \varepsilon_f) \quad (40)$$

$$i_t = \sigma(W_i X_{Lt} + \kappa_i \omega_{t-1} + \varepsilon_i) \quad (41)$$

$$o_t = \sigma(W_o X_{Lt} + \kappa_o \omega_{t-1} + \varepsilon_o) \quad (42)$$

$$\tilde{C}_t = \tanh\sigma(\kappa_c \omega_{t-1} + W_c X_{Lt} + \varepsilon_c) \quad (43)$$

$$\omega_t = o_t \otimes \tanh(i_t \otimes \tilde{C}_t + f_t \otimes C_{t-1}) \quad (44)$$

### D. The steps of SCKF-LSTM tracking method

The flows of the novel method are as follows:

**Algorithm 1** SCKF-LSTM based trajectory tracking algorithm
**Training the LSTM network using gas load data**
**Initialization:** $\hat{x}_0$, $z_1$, $u_1$, $S_{0|0}$
**repeat**
  *Prediction step*
  1) compute the predicted state $\tilde{x}_{t+1}$ and error covariance $\tilde{S}_{t+1}$ by (25)-(29)
  *Filtering step*
  2) compute the predicted measurement $\tilde{z}_{t+1}$ by (30)-(32)
  3) compute $S_{zz,t+1}$ and $P_{xz,t+1}$ by (33)-(36)
  4) compute the tracking state $\hat{x}_{t+1}$ and $\hat{S}_{t+1}$ by (37)-(39)
  **for** *i* **from** 1 **to** sink node number
    Predict the gas loads by the trained LSTM networks
  **end**
**until** *tracking time out*

### E. The computational complexity analysis

The calculation of SCKF includes the multiplications, square roots and inverses of matrixes, and the detailed complexities are shown in Tab. 2. The computational complexities of SCKF is $O(2/3n_x^3 + 7/3m_z^3 + 24n_x^2 + m_z^2 + 7n_x m_z + 3n_x + m_z)$. The variable $m_z$ is the measurement number. For the LSTM prediction, the computational complexities is $O(T \cdot L \cdot H^2)$. The variables $T$, $L$ and $H$ are the input sequence length, layer number and neuron number, respectively. The one step computational complexity of the proposed method is $O(2/3n_x^3 + 7/3m_z^3 + 24n_x^2 + m_z^2 + 7n_x m_z + 3n_x + m_z + T \cdot L \cdot H^2)$.

TABLE II
THE COMPUTATIONAL COMPLEXITIES OF THE METHOD

| variable | complexity | variable | complexity | variable | complexity |
|---|---|---|---|---|---|
| $\tilde{x}_{t+1}$ | $7n_x^2$ | $S_{zz,t+1}$ | $2n_x m_z + m_z^3/3$ | $\hat{x}_{t+1}$ | $2n_x + m_z$ |
| $\tilde{S}_{t+1}$ | $7n_x^2 + n_x^3/3$ | $P_{xz,t+1}$ | $4n_x^2$ | $\hat{S}_{t+1}$ | $2n_x^2 + n_x m_z + n_x + n_x^3/3$ |
| $\tilde{z}_{t+1}$ | $4n_x^2 + m_z^2 + 2n_x m_z$ | $W_{t+1}$ | $2m_z^2 + 2n_x m_z$ | | |

## IV. CASE STUDIES

### A. Testing System and LSTM Training

In the testing system, the IEEE 39-bus testing system in MATPOWER [34] and GasLib 40-node natural gas system [35] are coupled by 5 GTUs, which are given in Fig. 5. The gas sources are at node 1, 2 and 35, and the densities are constant. The blue trapezoids in the gas system are compressors. The density ratios of compressors are given in Tab. 3, while the specific details of the gas pipeline parameters are presented in Tab. 4.

TABLE III
RATIOS OF COMPRESSORS

| Compressors | 2-40 | 6-38 | 28-39 | 35-3 | 22-36 | 14-37 |
|---|---|---|---|---|---|---|
| Ratios | 1.2 | 1.25 | 1.1 | 1.2 | 1.15 | 1.18 |

First of all, the true values are generated by the simulations,

while the random errors are adding to true values to generate measurements. Then the novel tracking method is applied. In the simulation, the power flow results are obtained by the MATPOWER, and the nonlinear dynamic models of gas pipelines [36] applied by the specialized software for simulating natural gas systems, known as PipelineStudio [37] are used to generate the gas dynamic processes. In the proposed tracking method, the calculating interval is 5 minutes. The 30 day practical load data is collected from a gas station in Nanjing. The outliers and missing data are treated by linear interpolations. The proportions of training and testing sets for the LSTM are 80% and 10%, respectively, and the last 10% data is applied for the simulations. The load data $M_L$ is normalized by the following equation:

$$M_{LN} = (M_L - M_{Lmin})/(M_{Lmax} - M_{Lmin}) \quad (46)$$

The mean absolute percentage error (MAPE) serves as a metric to assess the accuracy of predictions. The MAPE values under different hyperparameters are given in Tab 5, and it can be seen that the smallest value is 0.2%. The input sequence length of the LSTM is 5, while the layer and hidden unit numbers are 3 and 80, respectively.

### B. Analysis of Simulation Result

The proposed tracking method is implemented on the constructed IGES, with specific tracking indices computed to

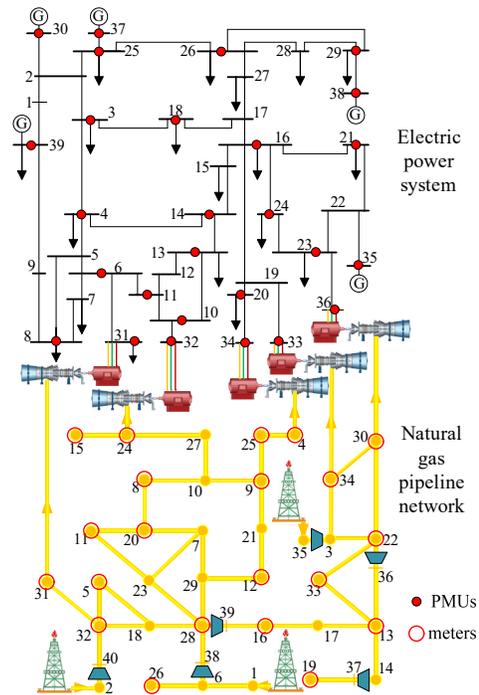

Fig. 5. The testing system.

### TABLE IV
### PARAMETERS OF GAS PIPELINES

| Pipe | Length (km) | Diameter (m) | Pipe | Length (km) | Diameter (m) | Pipe | Length (km) | Diameter (m) | Pipe | Length (km) | Diameter (m) | Pipe | Length (km) | Diameter (m) | Pipe | Length (km) | Diameter (m) |
|---|---|---|---|---|---|---|---|---|---|---|---|---|---|---|---|---|---|
| 1,6 | 13.07 | 1 | 12,21 | 10.02 | 0.6 | 10,27 | 38.65 | 0.4 | 20,11 | 10.45 | 0.6 | 32,5 | 31.17 | 0.8 | 30,22 | 26.42 | 0.8 |
| 14,19 | 76.89 | 0.8 | 29,7 | 35.21 | 0.6 | 25,4 | 18.01 | 0.6 | 6,26 | 12.39 | 0.8 | 5,18 | 12.76 | 1 | 13,14 | 18.13 | 1 |
| 28,16 | 21.55 | 1 | 7,23 | 20.32 | 0.6 | 27,24 | 3.06 | 0.6 | 11,23 | 19.30 | 0.6 | 32,2 | 32.92 | 0.8 | 13,22 | 65.05 | 0.8 |
| 16,17 | 6.99 | 1 | 21,9 | 32.86 | 0.8 | 24,15 | 12.01 | 0.4 | 28,23 | 66.03 | 0.6 | 3,22 | 49.86 | 0.8 | 13,33 | 65.53 | 1 |
| 17,13 | 58.21 | 0.8 | 28,6 | 47.48 | 0.8 | 10,8 | 14.04 | 0.4 | 28,18 | 18.96 | 1 | 22,33 | 3.47 | 0.8 | | | |
| 28,29 | 86.69 | 0.8 | 9,10 | 3.80 | 0.6 | 8,20 | 20.63 | 0.6 | 18,32 | 36.06 | 0.8 | 3,34 | 3.41 | 1 | | | |
| 29,12 | 16.57 | 0.6 | 9,25 | 39.03 | 0.8 | 20,7 | 10.58 | 0.6 | 32,31 | 22.22 | 0.8 | 30,34 | 32.44 | 1 | | | |

### TABLE V
### THE MAPE OF LSTM PREDICTION (%)

| | hidden units number | | | | | | | | |
|---|---|---|---|---|---|---|---|---|---|
| | 40 | | | 80 | | | 160 | | |
| Input length | 5 | 10 | 15 | 5 | 10 | 15 | 5 | 10 | 15 |
| 2 layers | 1.4 | 2.0 | 2.8 | 1.1 | 1.4 | 2.3 | 1.3 | 1.5 | 2.5 |
| 3 layers | 0.9 | 1.3 | 1.7 | **0.2** | 0.6 | 1.0 | 0.3 | 0.5 | 0.8 |

assess the method's performance. In the context of the power system, the tracking results for voltage phasor at bus 36 are illustrated in Fig. 6. In these representations, the actual values are illustrated with solid blue lines, whereas the red dashed lines correspond to the values being tracked. The close alignment of the tracking values with the true values is evident, underscoring the method's capability to accurately track dynamic trajectories with high precision.

The tracking results of gas states are presented in Fig.7. Similar to the findings in the power systems, the tracking performance for the gas system is highly accurate. The true values and tracking values of pressures are observed to nearly coincide. This high degree of accuracy can be attributed to the minimal variation in gas pressures and the precise predictions of gas densities within the system.

To study the performances numerically, the filter coefficients and the average variances [10] are computed. For power systems, the filter coefficients of voltages and currents are shown in Tab.6 and Tab.7, respectively. The filter coefficients of the buses without PMU or any load are not given in the tables. It can be seen that the filter coefficient values of

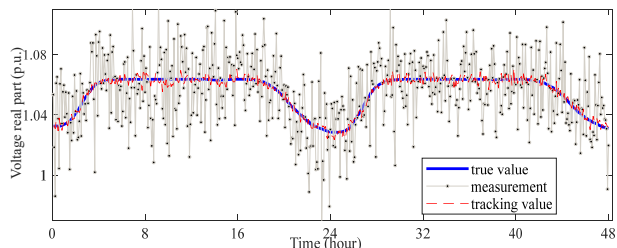
(a) the voltage real part of bus 36

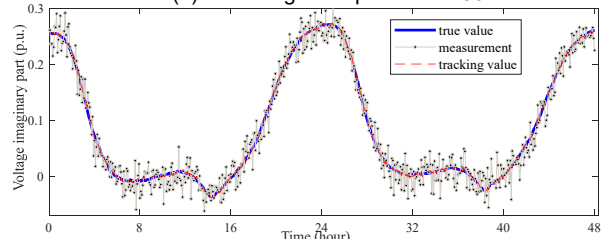
(b) the voltage imaginary part of bus 36

Fig. 6. Tracking results of voltage.

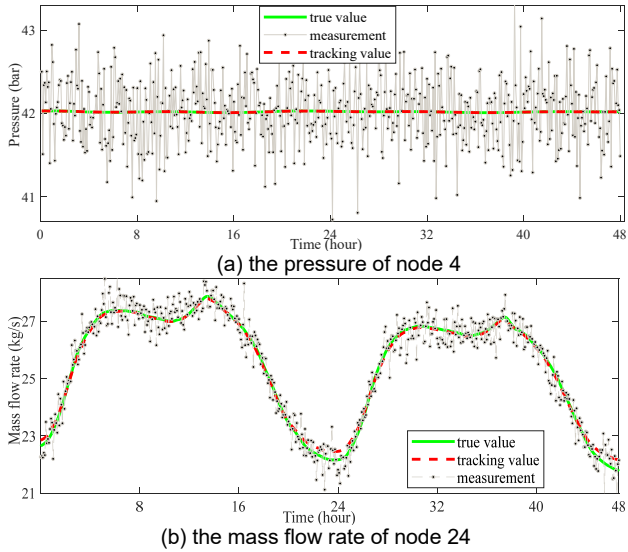
Fig. 7. Tracking results of gas states.

voltages are less than 1. This is attributed to the refinement of tracking values through prediction corrections and the effective filtering of random measurement errors. Consequently, the proposed tracking method succeeds in accurately obtaining dynamic trajectories for the system. Further insights are provided in Tab. 8 and 9, which display the tracking average variances for voltages and currents, respectively. It is important to note, however, that the average variances for currents at buses with zero injection are not included in these calculations.

In the context of gas systems, Tab. 10 presents the filtering coefficients for pressures and mass flow rates. The analysis omits nodes that lack measuring equipment, and it is observed that all filter coefficients are significantly less than 1. This finding is consistent with the results from the power systems and strongly indicates the proposed tracking method's capability to accurately track dynamic trajectories in gas systems. Furthermore, Tab. 11 presents the average variances for pressures and mass flow rates. The significantly low values of these variances confirm the efficacy of the proposed tracking method, further reinforcing its reliability and precision in tracking dynamic states within gas systems.

## C. Sensitivity Analysis and Robustness

The sensitivity analysis to measurement number and noises is given. The average values of filter coefficients with different sensor numbers and the tracking variances under different measurement variances are given in Tab. 12 and Tab.13, respectively. It shows that the higher the quantity and accuracy of the measurements, the higher the tracking precision.

The robustness of the method is discussed also. The 50% deviations are added to the measurements as bad data. The tracking results are given in Fig.8, showing that the tracking value fits the true value well under bad data condition.

TABLE VI
FILTER COEFFICIENTS RESULTS OF VOLTAGE PHASORS

| Bus | Real part | Imaginary part | Bus | Real part | Imaginary part | Bus | Real part | Imaginary part | Bus | Real part | Imaginary part | Bus | Real part | Imaginary part | Bus | Real part | Imaginary part |
|---|---|---|---|---|---|---|---|---|---|---|---|---|---|---|---|---|---|
| 3 | 0.0133 | 0.0147 | 8 | 0.0140 | 0.0154 | 13 | 0.0131 | 0.0139 | 18 | 0.0134 | 0.0148 | 23 | 0.0132 | 0.0141 | 26 | 0.0145 | 0.0158 |
| 4 | 0.0144 | 0.0142 | 10 | 0.0129 | 0.0147 | 14 | 0.0130 | 0.0145 | 20 | 0.0176 | 0.0158 | 24 | 0.0150 | 0.0147 | 29 | 0.0159 | 0.0149 |
| 6 | 0.0146 | 0.0144 | 11 | 0.0134 | 0.0152 | 16 | 0.0155 | 0.0163 | 21 | 0.0150 | 0.0144 | 25 | 0.0142 | 0.0148 | 30 | 0.0135 | 0.0172 |

TABLE VII
FILTER COEFFICIENTS OF CURRENTS

| Bus | Real part | Imaginary part | Bus | Real part | Imaginary part | Bus | Real part | Imaginary part | Bus | Real part | Imaginary part | Bus | Real part | Imaginary part |
|---|---|---|---|---|---|---|---|---|---|---|---|---|---|---|
| 3 | 0.8071 | 0.8281 | 20 | 0.5174 | 0.5495 | 26 | 0.8291 | 0.7771 | 33 | 0.4917 | 0.4215 | 38 | 0.4430 | 0.4080 |
| 4 | 0.9163 | 0.9145 | 21 | 0.9052 | 0.9038 | 29 | 0.6852 | 0.7226 | 34 | 0.4086 | 0.3864 | 39 | 0.5676 | 0.5981 |
| 8 | 0.8101 | 0.7727 | 23 | 0.7798 | 0.8255 | 30 | 0.4793 | 0.4163 | 35 | 0.4341 | 0.4980 | | | |
| 16 | 0.8818 | 0.8758 | 24 | 0.8536 | 0.8360 | 31 | 0.4052 | 0.4026 | 36 | 0.4721 | 0.4282 | | | |
| 18 | 0.5745 | 0.6209 | 25 | 0.6973 | 0.6975 | 32 | 0.4109 | 0.4151 | 37 | 0.3742 | 0.4152 | | | |

TABLE VIII
AVERAGE VARIANCE RESULTS OF VOLTAGE PHSORS ($10^{-5}$)

| Bus | Real part | Imaginary part | Bus | Real part | Imaginary part | Bus | Real part | Imaginary part | Bus | Real part | Imaginary part | Bus | Real part | Imaginary part | Bus | Real part | Imaginary part |
|---|---|---|---|---|---|---|---|---|---|---|---|---|---|---|---|---|---|
| 1 | 0.6150 | 0.6323 | 8 | 0.5612 | 0.6031 | 15 | 0.5539 | 0.5850 | 22 | 0.5668 | 0.5877 | 29 | 0.6509 | 0.6508 | 36 | 0.5882 | 0.6186 |
| 2 | 0.5725 | 0.5967 | 9 | 0.5938 | 0.6395 | 16 | 0.5554 | 0.5824 | 23 | 0.5665 | 0.5909 | 30 | 0.5828 | 0.6118 | 37 | 0.5732 | 0.6117 |
| 3 | 0.5597 | 0.5891 | 10 | 0.5585 | 0.5942 | 17 | 0.5575 | 0.5889 | 24 | 0.5572 | 0.5812 | 31 | 0.5600 | 0.6194 | 38 | 0.6602 | 0.6605 |
| 4 | 0.5607 | 0.5921 | 11 | 0.5571 | 0.5940 | 18 | 0.5573 | 0.5915 | 25 | 0.5709 | 0.5978 | 32 | 0.5752 | 0.6148 | 39 | 0.6186 | 0.6507 |
| 5 | 0.5575 | 0.5998 | 12 | 0.5701 | 0.6063 | 19 | 0.5780 | 0.6058 | 26 | 0.5859 | 0.6107 | 33 | 0.5874 | 0.6142 | | | |
| 6 | 0.5568 | 0.5995 | 13 | 0.5566 | 0.5922 | 20 | 0.5950 | 0.6242 | 27 | 0.5629 | 0.5999 | 34 | 0.6121 | 0.6411 | | | |
| 7 | 0.5588 | 0.6025 | 14 | 0.5558 | 0.5912 | 21 | 0.5581 | 0.5829 | 28 | 0.6502 | 0.6389 | 35 | 0.5698 | 0.5916 | | | |

TABLE XI
AVERAGE VARIANCES OF CURRENTS ($10^{-5}$)

| Bus | Real part | Imaginary part | Bus | Real part | Imaginary part | Bus | Real part | Imaginary part | Bus | Real part | Imaginary part | Bus | Real part | Imaginary part |
|---|---|---|---|---|---|---|---|---|---|---|---|---|---|---|
| 1 | 0.2834 | 0.2642 | 12 | 0.3921 | 0.3401 | 23 | 0.2931 | 0.3370 | 29 | 0.2830 | 0.2794 | 35 | 0.1796 | 0.1863 |
| 3 | 0.3219 | 0.3336 | 15 | 0.3366 | 0.3750 | 24 | 0.3533 | 0.3180 | 30 | 0.1925 | 0.1741 | 36 | 0.1752 | 0.1646 |
| 4 | 0.3828 | 0.3434 | 16 | 0.3762 | 0.3721 | 25 | 0.2619 | 0.2391 | 31 | 0.1688 | 0.1748 | 37 | 0.1591 | 0.1541 |
| 7 | 0.2684 | 0.2926 | 18 | 0.2559 | 0.2428 | 26 | 0.3021 | 0.3006 | 32 | 0.1724 | 0.1666 | 38 | 0.1873 | 0.1831 |
| 8 | 0.3365 | 0.3401 | 20 | 0.2085 | 0.2232 | 27 | 0.2951 | 0.2717 | 33 | 0.1979 | 0.1662 | 39 | 0.2292 | 0.2509 |
| 9 | 0.2405 | 0.2201 | 21 | 0.3435 | 0.3738 | 28 | 0.2788 | 0.3064 | 34 | 0.1637 | 0.1481 | | | |

TABLE X
FILTER COEFFICIENTS OF GAS STATES

| node | Pressures ($10^{-6}$) | Gas Load ($10^{-10}$) | node | Pressures ($10^{-6}$) | Gas Load ($10^{-10}$) | node | Pressures ($10^{-6}$) | Gas Load ($10^{-10}$) | node | Pressures ($10^{-6}$) | Gas Load ($10^{-10}$) |
|---|---|---|---|---|---|---|---|---|---|---|---|
| 4 | 0.5633 | 2.5914 | 11 | 0.2513 | 1.5495 | 22 | 0.0059 | 6.2340 | 30 | 0.0096 | 4.6021 |
| 5 | 0.0249 | 1.5558 | 15 | 0.9346 | 0.3892 | 24 | 0.5562 | 3.7290 | 31 | 0.0610 | 7.4352 |
| 8 | 0.2676 | 1.5612 | 16 | 0.0191 | 1.5618 | 25 | 0.5927 | 1.5234 | 33 | 0.0071 | 1.5509 |
| 9 | 0.6117 | 3.5090 | 19 | 0.0277 | 0.3889 | 26 | 0.0058 | 3.8932 | 34 | 0.0064 | 1.7352 |
| 10 | 0.5387 | 3.4857 | 20 | 0.2185 | 3.4948 | 28 | 0.0019 | 9.7602 | | | |

TABLE XI
AVERAGE VARIANCES OF GAS STATES

| node | Pressures ($10^{-8}$) | Gas Load ($10^{-11}$) | node | Pressures ($10^{-8}$) | Gas Load ($10^{-11}$) | node | Pressures ($10^{-8}$) | Gas Load ($10^{-11}$) | node | Pressures ($10^{-8}$) | Gas Load ($10^{-11}$) |
|---|---|---|---|---|---|---|---|---|---|---|---|
| 4 | 9.5069 | 1.3452 | 12 | 6.2847 | 2.3728 | 20 | 3.8142 | 5.7281 | 28 | 0.0308 | 9.9270 |
| 5 | 0.3632 | 2.6285 | 13 | 0.1633 | 9.7630 | 21 | 8.1529 | 2.7029 | 29 | 0.5063 | 5.1221 |
| 6 | 0.0493 | 5.1324 | 14 | 0.1925 | 2.4167 | 22 | 0.1026 | 9.4930 | 30 | 0.1473 | 2.9374 |
| 7 | 3.6662 | 5.1329 | 15 | 7.2173 | 0.5185 | 23 | 3.4026 | 5.9590 | 31 | 0.9060 | 2.4693 |
| 8 | 4.6009 | 2.4084 | 16 | 0.3070 | 2.5816 | 24 | 7.1407 | 2.2388 | 32 | 0.3583 | 8.9628 |
| 9 | 9.1896 | 5.5641 | 17 | 0.3031 | 2.4084 | 25 | 9.3241 | 2.5886 | 33 | 0.1113 | 2.3749 |
| 10 | 9.1389 | 5.7958 | 18 | 0.3359 | 5.8363 | 26 | 0.0104 | 0.5430 | 34 | 0.0010 | 7.9586 |
| 11 | 3.7372 | 2.7123 | 19 | 0.4561 | 0.5968 | 27 | 7.0572 | 2.2605 | | | |

TABLE XII
THE FILTER COEFFICIENTS WITH DIFFERENT MEASUREMENT NUMBERS

| Sensor Number | Pressure | Mass | e | f |
|---|---|---|---|---|
| 33 PMUs, 26 meters | $0.24 \times 10^{-6}$ | $0.30 \times 10^{-9}$ | 0.009 | 0.009 |
| 27 PMUs, 21 meters | $0.28 \times 10^{-6}$ | $0.32 \times 10^{-9}$ | 0.01 | 0.01 |
| 24 PMUs, 19 meters | $0.78 \times 10^{-6}$ | $1.49 \times 10^{-9}$ | 0.043 | 0.044 |

TABLE XIII
THE TRACKING RESULTS WITH DIFFERENT MEASUREMENT VARIANCES

| Measurement variance($10^{-4}$) | | | | Tracking variance | | | |
|---|---|---|---|---|---|---|---|
| voltage | current | pressure | mass | Pressure | Mass | e | f |
| 9 | 9 | 4 | 9 | $5.29 \times 10^{-8}$ | $8.57 \times 10^{-11}$ | $0.72 \times 10^{-5}$ | $1.09 \times 10^{-5}$ |
| 4 | 4 | 1 | 4 | $3.11 \times 10^{-8}$ | $4.14 \times 10^{-11}$ | $0.57 \times 10^{-5}$ | $0.61 \times 10^{-5}$ |
| 1 | 1 | 0.64 | 1 | $2.01 \times 10^{-8}$ | $1.95 \times 10^{-11}$ | $0.21 \times 10^{-5}$ | $0.20 \times 10^{-5}$ |

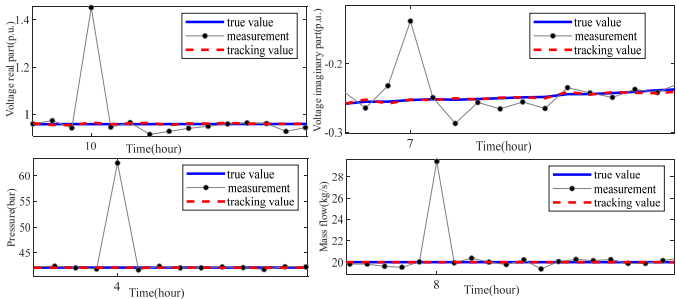

Fig. 8. The tracking results under bad data condition.

### D. Comparison with Other Method

The performance of SCKF-LSTM is evaluated by comparison with that obtained by linear method in [11] and the multi-layer perceptron (MLP) based prediction method. The three algorithms are simulated 200 times, and the average values of filter coefficients are computed, which are shown in Tab. 14. It is clear that the values obtained from the SCKF-LSTM are consistently lower than those reported in reference [11] and SCKF-MLP, meaning that the proposed method stands out as the most precise. In particular, the value of mass flow rate of [11] is large due to errors in the gas model arise caused the simplification of the gas dynamics equations through linearization.

TABLE XIV
THE COMPARISON RESULTS OF FILTER COEFFICIENTS

| | Pressure | Mass | e | f |
|---|---|---|---|---|
| SCKF-LSTM | $0.28 \times 10^{-6}$ | $0.32 \times 10^{-9}$ | 0.01 | 0.01 |
| SCKF-MLP | $0.57 \times 10^{-5}$ | $0.81 \times 10^{-7}$ | 0.03 | 0.03 |
| [11] | $0.88 \times 10^{-5}$ | 0.41 | 0.04 | 0.24 |

### E. Discussion of the Observability

For the purpose of examining the observability of the method, an matrix $Q_B$ is formulated as follows:

$$Q_B = \begin{bmatrix} C \\ C\Phi \\ \vdots \\ C\Phi^{n_x - 1} \end{bmatrix} \quad (46)$$

According to Kalman criterion, if the rank of matrix $Q_B$ is $n_x$, the system is observable.

In the tracking model, the total number of states is 196, and the rank of $Q_B$ is 196 under the sensor equipment as Fig. 5, meaning the system is observable.

## V. CONCLUSION

This paper introduces a SCKF-LSTM based trajectory tracking method for IESs, aimed at capturing accurate dynamic processes. The method utilizes the Euler method for discretizing the PDEs of gas systems, while Holt's exponential smoothing technology forms the basis of the power system equations. The gas loads are predicted using the LSTM network, followed by the SCKF solving the tracking problem. In the conducted case studies, an IES is modeled to evaluate the efficacy of this innovative approach. The simulations confirm the method's precision in tracking dynamic trajectories of IESs. Moreover, for a quantitative assessment of tracking proficiency, metrics including filter coefficients and mean variances have been applied. The findings reveal that all filter coefficient values are below 1, indicating that the tracking precision surpasses that of the measurements.